\documentclass[twocolumn,showpacs,preprintnumbers,amsmath,amssymb,prl,superscriptaddress]{revtex4}

\usepackage{graphicx}
\usepackage{dcolumn}
\usepackage{bm}

\begin{document}

\title{Vanishing Hall Coefficient in the Extreme Quantum Limit in Photocarrier-Doped SrTiO$_{3}$}

\author{Y. Kozuka}
\altaffiliation{kk077105@mail.ecc.u-tokyo.ac.jp}
\affiliation{Department of Advanced Materials Science, University of Tokyo, Kashiwa, Chiba 277-8561, Japan}

\author{T. Susaki}
\affiliation{Department of Advanced Materials Science, University of Tokyo, Kashiwa, Chiba 277-8561, Japan}

\author{H. Y. Hwang}
\affiliation{Department of Advanced Materials Science, University of Tokyo, Kashiwa, Chiba 277-8561, Japan}
\affiliation{Japan Science and Technology Agency, Kawaguchi, 332-0012, Japan}

\begin{abstract}
We have investigated the extreme quantum limit of photogenerated electrons in quantum paraelectric SrTiO$_{3}$. This regime is distinct from conventional semiconductors, due to the large electron effective mass and large lattice dielectric constant. At low temperature, the magnetoresistance and Hall resistivity saturate at high magnetic field, deviating from conventional behavior. As a result, the Hall coefficient vanishes on the scale of the ratio of the Landau level splitting to the thermal energy, indicating the essential role of lowest Landau level occupancy, as limited by thermal broadening.
\end{abstract}

\pacs{72.20.My, 72.40.+w, 71.70.Di, 77.84.Bw}

\maketitle

Magnetotransport properties in the extreme quantum limit \cite{Roth}, where only the lowest Landau level is partially occupied, has been an attractive topic in semiconductor physics as a natural extension of the Shubnikov-de Haas effect in three dimensions \cite{Halperin} and integer quantum Hall effect in two dimensions \cite{Tsui}. The requirements to reach the quantum limit are \cite{Roth}
\begin{eqnarray}
\omega_{c}\tau &>& 1,\\
\hbar\omega_{c} &>& E_{F},\ kT,
\end{eqnarray}
where $\omega_{c}=eH/m^{*}$ is the cyclotron frequency ($e$ is the elementary charge, $H$ is the magnetic field, and $m^{*}$ is the electron effective mass),  $\tau$ is the carrier relaxation time, $E_{F}$ is the Fermi energy, and $kT$ is the thermal energy. In bulk semiconductors, there are few candidates to reach the quantum limit because a carrier density $n<10^{17}$ cm$^{-3}$ is needed, based on condition (2) within a free carrier approximation under a typical laboratory magnetic field of 10 T. Since most semiconductors are insulating in the range of such low $n$, it is difficult to maintain sufficiently high electron mobility to satisfy condition (1). According to the Mott criterion, the critical carrier density for a metal-insulator transition is given by $n_{c} \simeq (0.25/a^{*})^{3}$, where $a^{*}=\epsilon\hbar^{2}/m^{\ast}e^{2}$ is the effective Bohr radius ($\epsilon$ is the dielectric constant) \cite{Mott}. Applying the Mott criterion to condition (2), we find a necessary condition as $\epsilon_{r}/m_{r} > 1.3 \times 10^{3}$, where $\epsilon_{r}$ and $m_{r}$ are defined by $\epsilon/\epsilon_{0}$ ($\epsilon_{0}$ is the vacuum permittivity) and $m^{\ast}/m_{0}$ ($m_{0}$ is the bare electron mass), respectively. In terms of this requirement, the quantum limit has been extensively studied in narrow-gap semiconductors such as InSb ($m^{\ast}=0.014m_{0}$, $\epsilon_{r}=16$) and Hg$_{1-x}$Cd$_{x}$Te ($x\simeq 0.2$, $m^{\ast}=0.007m_{0}$, $\epsilon_{r}=17$) because of their extremely small effective mass \cite{Shayegan}. In this limit, magnetic freeze-out has usually been observed, in which the magnetic field induces the localization of carriers by shrinking the electron wavefunction.  Wigner crystallization has also been proposed as the high-field ground state \cite{Nimtz,Rosenbaum}.\par
Quantum paraelectric materials such as SrTiO$_{3}$ can provide alternative materials candidates due to the large lattice dielectric constant, exceeding 20,000 at low temperature \cite{Sakudo}. In addition, electrons can have rather high Hall mobility $\mu$ ($>10,000$ cm$^{2}$ V$^{-1}$ s$^{-1}$), easily satisfying condition (1) \cite{Tufte}. However, the large effective mass ($m^{\ast}=1.2m_{0}$ \cite{Uwe1}) dictates that the approach to the quantum limit is rather different than for narrow-gap semiconductors. In those systems with small $m^{\ast}$, intermediate temperatures are sufficient to meet condition (1) because of the large Landau level splitting. Similarly, at low temperatures $E_{F}$ can also be relatively large, so that the starting point under zero magnetic field is a degenerate electron gas. With increasing magnetic field, nondegeneracy is caused by the lowering of $E_{F}$ following $E_{F}\propto 1/H^{2}$ \cite{Rosenbaum}. By contrast, for SrTiO$_{3}$, the small Landau level splitting requires low temperature and low $E_{F}$. At such low densities, the experimentally accessible zero field state is a nondegenerate electron gas. Thus, the regime of the quantum limit obtained in SrTiO$_{3}$ is rather unique, and has never been experimentally studied.\par
In this Letter, we report the results of magnetotransport studies of SrTiO$_{3}$ in the nondegenerate quantum limit down to 2 K. We have used photocarrier doping to create a homogeneous, low carrier density electron gas far below that accessible by chemical substitution. Chemical doping is further limited by a tendency toward intrinsic clustering, as observed for oxygen vacancies \cite{Szot,Cuong,Muller}. Although the low-field magnetotransport properties exhibited conventional semiconductor behavior, as previously reported \cite{Yasunaga,Feng,Kozuka}, at high magnetic fields the transverse magnetoresistance and Hall resistivity saturate. This results in a strong drop in the magnetic field dependent Hall coefficient, which was well scaled by $\hbar\omega_{c}/kT$, indicating a tendency toward a vanishing Hall coefficient for the partially occupied lowest Landau level, dominantly limited by thermal effects. This scaling was found to be robust over almost 2 decades in carrier density.\par
SrTiO$_{3}$ single crystals were cut into typical dimensions of $3.0\times0.5\times0.5$ mm$^{3}$, and electrical contacts were made by In ultrasonic soldering at the sides of the sample. The experimental details of the photoconductivity measurements have been given previously \cite{Kozuka}. The light intensity and wavelength were fixed to constant values of $\sim$ 0.05 mW/cm$^{2}$ (unless otherwise indicated) and $\lambda=380$ nm. At this wavelength, just at the band edge, the carrier density profile perpendicular to the sample surface can be neglected because of the large optical penetration depth, far exceeding the thickness of the crystal (the light intensity is estimated to vary by $<10$ \% through the thickness of the sample) \cite{Capizzi}. In these conditions, the photoconductivity of SrTiO$_{3}$ has been well established. The photogenerated carriers thermalize on the time scale of $\hbar/kT=4\times 10^{-12}$ s (2 K), while the photoluminescent decay (roughly the carrier lifetime) is $\sim1$ ms at low temperatures \cite{Hasegawa}. Thus under continuous illumination, a steady-state quasi-equilibrium is achieved. The mobile carriers are electrons, the holes are trapped, and the low-field magnetotransport coefficients are typical of electron-doped SrTiO$_{3}$ \cite{Tufte,Yasunaga,Feng,Kozuka}. The temperature dependence of the low-field transport properties (0 T data) for these studies are given in Fig. 1 (a)-(c). To summarize the basic parameters of the photogenerated electrons under these illumination conditions, a 3D electron gas is formed with $n(\mathrm{2\ K})=2 \times 10^{13}$ cm$^{-3}$, $E_{F}/k=27$ mK, and $\hbar/\tau (\mathrm{2\ K})=0.8$ K.\par
\begin{figure}[tbp]
  \begin{center}
    \includegraphics{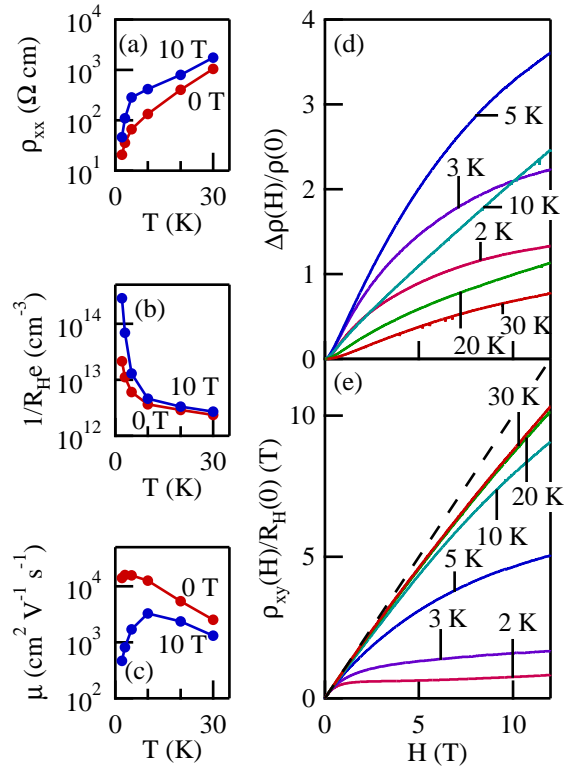}
  \end{center}
  \caption{(Color online) Temperature dependence of (a) resistivity $\rho_{xx}$, (b) $1/R_{H}e$, and (c) mobility $\mu$ of photocarrier-doped SrTiO$_{3}$ at 0 T and 10 T under continuous illumination at $\lambda=380$ nm, 0.05 mW/cm$^{2}$. (d) Transverse magnetoresistance and (e) Hall resistivity normalized by  the low-field Hall coefficient as a function of magnetic field.}
\end{figure}
Figure 1 (d) and (e) shows the transverse magnetoresistance and Hall resistivity $\rho_{xy}$ of SrTiO$_{3}$ (100) as a function of magnetic field with varying temperature. $\rho_{xy}$ is normalized by the Hall coefficient $R_{H}=\rho_{xy}/H$ in the low-field limit. At low field, $\rho_{xx}$ and $\rho_{xy}$ are well described by conventional expressions, $\rho_{xx}=\zeta(\mu H)^2$ ($\zeta$ is a constant of order one) and $\rho_{xy}=H/ne$, respectively. At high fields however, particularly with decreasing temperature, $\rho_{xy}$ saturates. We confirmed that this nonlinearity is also present in the cases of $H||$(110) and $H||$(111) by using different orientation samples, ruling out anisotropic effects. To further examine this unusual feature, the normalized magnetic field dependent Hall coefficient $R_{H}(H)$ is given in Fig. 2 (a). $R_{H}$ is almost constant at higher temperature, while it rapidly decreases by an order of magnitude by 2 K.\par
A number of scenarios for this drop in $R_{H}$ within a Boltzmann transport picture can be ruled out. A real increase in $n$ induced by magnetic field would imply a dramatic drop in $\mu$ [10 T data in Fig. 1 (a)-(c)], highly inconsistent with the negligible doping dependence observed in this range of density \cite{Kozuka}. In spite of the resemblance to the magnetic anomalous Hall effect \cite{Berger}, our results imply a loss of scattering at high fields, not additional spin scattering at low fields. Similarly, we can exclude the explanation of multi-band conduction or magnetic breakdown, because $R_{H}=1/n_{eff}e$ is a robust result in the high-field limit, where $n_{eff}$ is potentially a net compensated carrier density \cite{Berger}.\par
\begin{figure}[tbp]
  \begin{center}
    \includegraphics{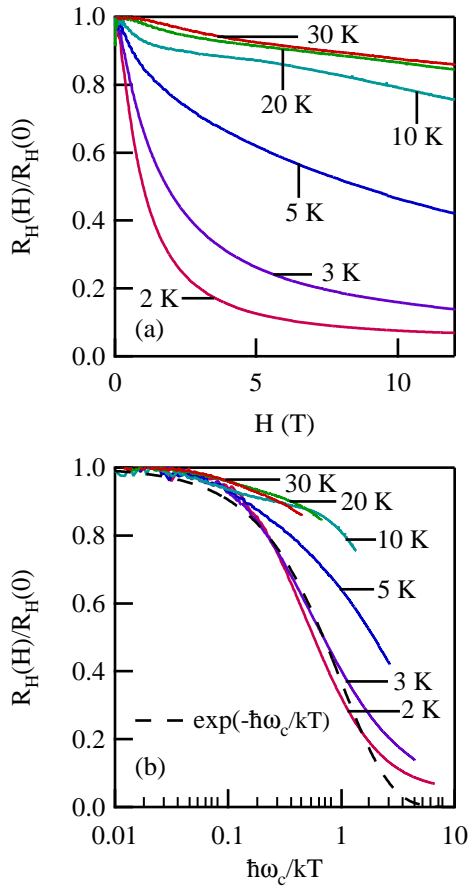}
  \end{center}
  \caption{(Color online) Hall coefficient of photocarrier doped SrTiO$_{3}$ ($\lambda=380$ nm, 0.05 mW/cm$^{2}$), normalized by its low-field value as a function of (a) magnetic field and (b) $\hbar\omega_{c}/kT$. The dashed line in (b) indicates the Boltzmann function.}
\end{figure}
To understand what determines the magnetic field scale for the saturation in $\rho_{xy}$, it is helpful to consider the hierarchy of the relevant energies.  Since $E_{F}$ is so small, the nondegenerate electrons have $kT$ energy.  This is also larger than scattering effects for all of these measurements ($kT>\hbar/\tau$), given the high electron mobility ($\omega_{c}\tau = 17$ at 2 K at the maximum field of 12 T, using the low-field transport coefficients).  In this case, the nondegenerate quantum limit is determined by the ratio $\hbar\omega_{c}/kT$, which gives the scale for having well-defined lowest Landau level occupancy.  Figure 2 (b) shows that the loss of $R_{H}$ is set by this scale.\par
\begin{figure}[tbp]
  \begin{center}
    \includegraphics{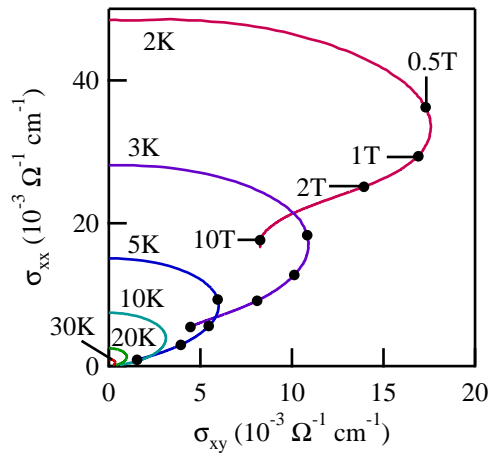}
  \end{center}
  \caption{(Color online) $\sigma_{xx}$ as a function of $\sigma_{xy}$ for photocarrier doped SrTiO$_{3}$ ($\lambda=380$ nm, 0.05 mW/cm$^{2}$). The points corresponding to H=0.5 T, 1 T, 2 T, and 10 T are shown for the data at 5 K and below.}
\end{figure}
Perhaps the phenomenologically closest observation to these results is the $\sim 10-40$ \% reduction of $R_{H}$ in narrow-gap semiconductors, just prior to to the onset of magnetic freeze-out (sometimes referred to as the ``Hall dip'' or ``Hall plateau'') \cite{Goldman,Mani,Viehweger}. Scenarios proposed to explain this feature include inhomogeneous metallic clusters \cite{Goldman}, three-dimensional quantum Hall effect \cite{Mani}, and Zeeman splitting of the lowest Landau level \cite{Viehweger}. Although these theories predict $R_{H}$ reduction around the critical field for magnetic freeze-out, strong divergence of $\rho_{xx}$ and $\rho_{xy}$ has been experimentally observed in the extreme quantum limit, which is absent in our measurements. The form of this divergence has been argued by some to suggest a Wigner crystal \cite{Nimtz,Rosenbaum}. One of the key parameters for magnetic freeze-out is the binding energy $E_{B}$, which is strongly affected by $m^{\ast}$, similar to the other energies discussed above. $E_{B}$ in the high field limit can be approximated as \cite{Shklovkii}
\begin{equation}
E_{B}(H)\simeq \frac{\hbar^{2}}{2m^{*}a^{*}}\left[\ln\left(\frac{a^{*}}{l}\right)^{2}\right]^{2},
\end{equation}
where $l=\sqrt{\hbar/eH}$ is the magnetic length. Even at 10 T the binding energy corresponds to $\simeq 1.4$ K for SrTiO$_{3}$. The large $m^{\ast}$ pushed the state of magnetic freeze-out to the regime of extremely low temperature and high magnetic field, and thus would appear unrelated to the saturation in $\rho_{xy}$ observed here.\par
We also note that the form suggested for partial freeze-out in magnetic clusters \cite{Goldman} does not describe our data. Instead, the simplest approximation is the Boltzmann function, as shown in Fig. 2 (b).  Although the significance of this is presently unclear, it is striking that the crossover in $R_{H}$ to the extreme quantum limit follows a classical form. A related point of comparison are studies above 50 K in conventional semiconductors of approaching the quantum limit starting from nondegenerate electrons ($kT>E_{F}$) \cite{Bate}. There only a weak feature ($<20$ \% change) was observed crossing $\hbar\omega_{c}/kT\sim 1$.  In those higher temperature studies, acoustic phonon or piezoelectric scattering was dominant, whereas our experiments at low temperature in SrTiO$_{3}$ were dominated by ionized impurity scattering (as well as the trapped holes).
\par
\begin{figure}[tbp]
  \begin{center}
    \includegraphics{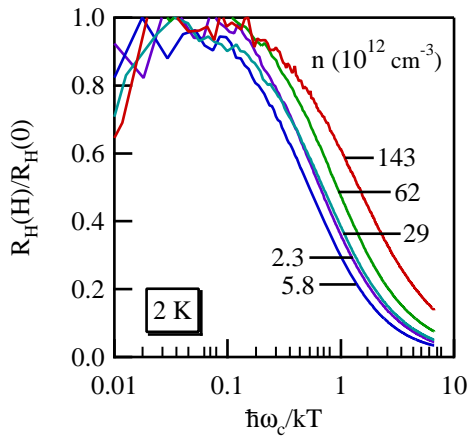}
  \end{center}
  \caption{(Color online) Hall coefficient for photocarrier doped SrTiO$_{3}$ ($\lambda=380$ nm) normalized by its low-field value as a function of $\hbar\omega_{c}/kT$, for varying carrier density obtained using $0.005-0.5$ mW/cm$^{2}$ light intensity at 2 K.}
\end{figure}
Another perspective on the unusual high field transport properties can be seen in Fig. 3, where $\sigma_{xx}$ is given as a function of $\sigma_{xy}$ at various temperatures.  At high temperature, the conventional parabolic relationship is observed.  In particular, with $\rho_{xx}$ saturated, the high field approach to the origin is given by $\sigma_{xx}\propto\sigma_{xy}^{2}$.  With decreasing temperature, however, increasing deviations from this form are observed, such that by 2 K, it is difficult to imagine a conventional extrapolation for $H\to\infty$.  Given that we can rule out virtually all conventional mechanisms as discussed above, we are left to speculate whether a collective state is approached at high field.  This would be unexpected in the following sense.  The electron energy ($kT \simeq  0.09 \times T$ meV) is lower than the energy of the soft polar optical phonon ($\hbar\omega_{phonon}(\mathrm{2\ K}) \simeq  2$ meV \cite{Yamada}), which is the phonon approaching a soft-mode transition that classically would lead to ferroelectricity.  The carriers can be well dynamically screened by the lattice despite poor electron-electron screening at low $n$, in addition to static screening of impurities or trapped holes. Thus we expected to observe a conventional high-field limit for weakly interacting electrons, which is not what is experimentally found.\par
Finally, we have extended these studies over a broad range of carrier densities by varying the intensity of illumination at $\lambda=380$ nm.  All of the features were quite similar to those given in Figs. 1-3. To summarize the central observation of the vanishing Hall coefficient, Fig. 4 gives the scaled $R_{H}$ for data spanning a factor of 62 change in density. In all cases, the data are well described by a drop when $\hbar\omega_{c}/kT\sim 1$.\par
In conclusion, we have demonstrated that quantum paraelectric materials can reach the quantum limit due to their large dielectric constant, utilizing homogeneous, low carrier density doping by photocarrier generation. Due to the large effective mass and small Fermi energy, we have studied the low-temperature nondegenerate quantum limit, a regime which has never been previously investigated. The magnetoresistance and Hall resistivity in SrTiO$_{3}$ were observed to saturate at high magnetic field, and the loss of the Hall coefficient is well described by the parameter, $\hbar\omega_{c}/kT$. Although a detailed analysis is not accessible due to the lack of a formal theory, the occupancy of the lowest Landau level is found to be crucial for these observations. Although these data are limited to 2 K by our current experimental capabilities, we note that the extremely low light intensities needed allow, in principle, access to dilution refrigerator temperatures.  Lower temperature studies should shed further light on the effects of magnetic freeze-out and disorder on these phenomena.\par
We thank N. Nagaosa, J. Zaanen, M. Onoda, S. Onoda, and K. S. Takahashi for helpful discussions.

\end{document}